\documentclass{webofc}
\usepackage[varg]{txfonts}   % Web of Conferences font
\usepackage{hyperref}
\usepackage{url}
\usepackage{lineno}
\hypersetup{colorlinks=true,citecolor=blue,urlcolor=blue,linkcolor=blue}
\usepackage{graphicx}% Include figure files

\newcommand{\Fermi}{{\textit{Fermi}}}

\newcommand{\size}{4.6}

\newcommand{\be}{\begin{equation}}
\newcommand{\ee}{\end{equation}}
\newcommand{\bi}{\begin{itemize}}
\newcommand{\ei}{\end{itemize}}
\newcommand{\ben}{\begin{enumerate}}
\newcommand{\een}{\end{enumerate}}
\begin{document}

\title{Classification of \Fermi-LAT unassociated sources with machine learning in the presence of dataset shifts}

\author{\firstname{Dmitry V.} \lastname{Malyshev}\inst{1}\fnsep\thanks{\email{dmitry.malyshev@fau.de}} 
}

\institute{
Erlangen Centre for Astroparticle Physics, Nikolaus-Fiebiger-Str. 2, Erlangen 91058, Germany
          }

\abstract{
About one third of {\Fermi} Large Area Telescope (LAT) sources are unassociated. We perform multi-class classification of \Fermi-LAT sources using machine learning with the goal of probabilistic classification of the unassociated sources. A particular attention is paid to the fact that the distributions of associated and unassociated sources are different as functions of source parameters. In this work, we address this problem in the framework of dataset shifts in machine learning.
}
\maketitle

%
%\vspace{0mm}

\section{Dataset shifts}
\label{sect:shifts}

The basic assumption of classification with machine learning is that the joint distributions of input features $x$
and output features, i.e., classes $k$, are the same for the training and target datasets:
\be
p_{\rm train} (x, k) = p_{\rm target} (x, k).
\ee
In the presence of a dataset shift the training and target distributions are different
$p_{\rm train} (x, k) \neq p_{\rm target} (x, k)$.
The joint distribution can be written as a product of conditional probability times a prior distribution in two different ways:
\be
p(x, k) = p(k|x) p(x) = p(x|k) p(k).
\ee
Correspondingly, there are two special cases of the dataset shift \citep{MorenoTorres2012AUV}:
\vspace{-1mm}
\ben
\item
Covariate shift: $p_{\rm train}(k|x) = p_{\rm target}(k|x)$, but $p_{\rm train}(x) \neq p_{\rm target}(x)$;
\vspace{-1mm}
\item
Prior shift: $p_{\rm train}(x|k) = p_{\rm target}(x|k)$, but $p_{\rm train}(k) \neq p_{\rm target}(k)$.
\een
\vspace{-1mm}

\begin{figure}[ht]
\centering
\vspace{-3mm}
\includegraphics[width={\size}cm,clip]{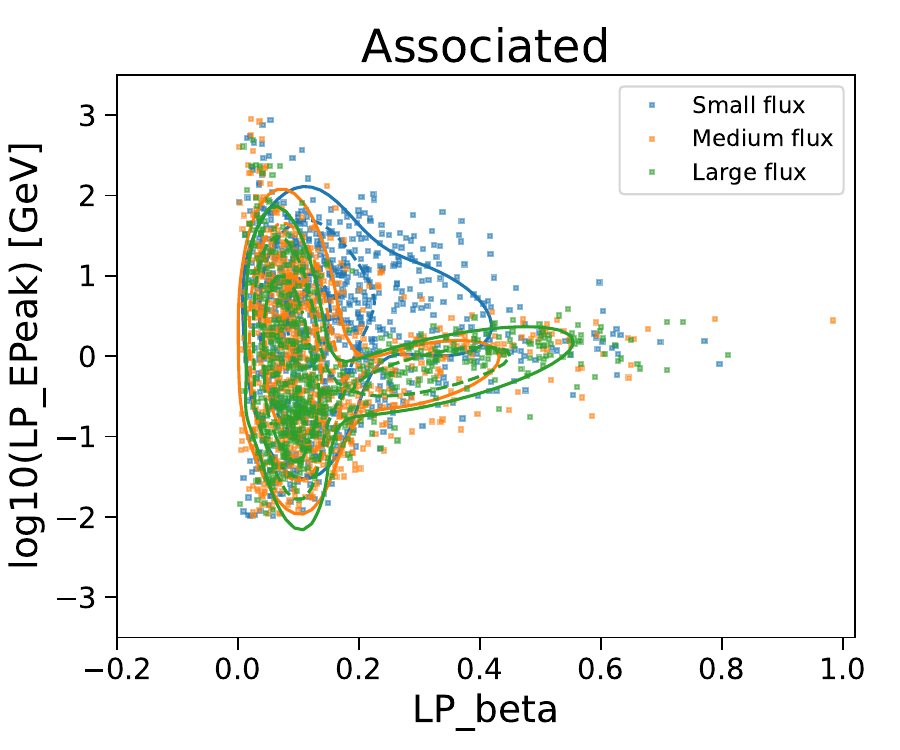}
\includegraphics[width={\size}cm,clip]{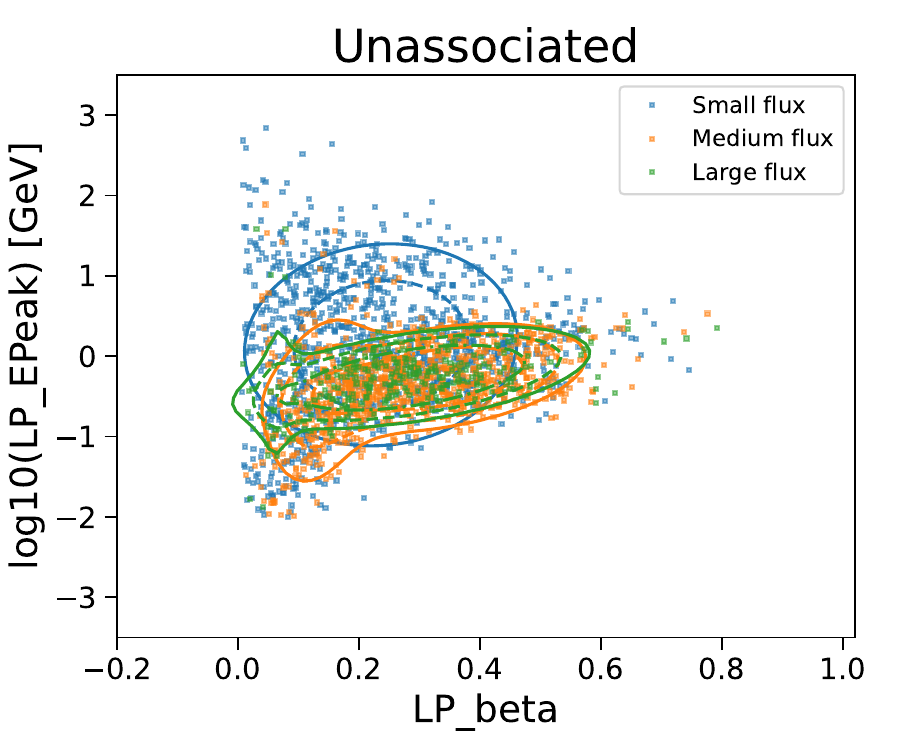} \\
\includegraphics[width={\size}cm,clip]{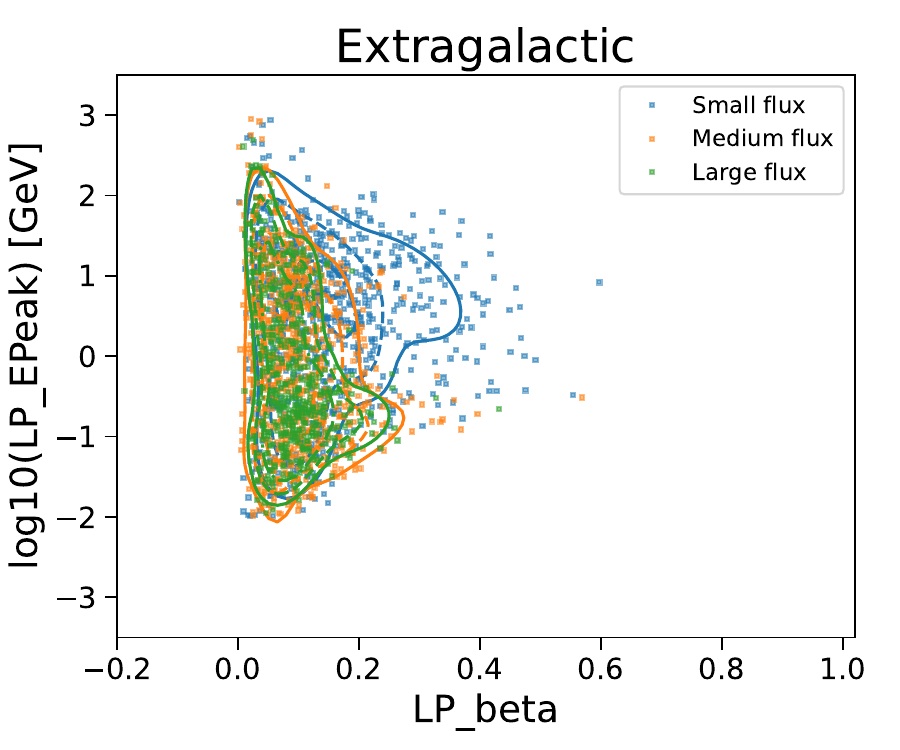}
\includegraphics[width={\size}cm,clip]{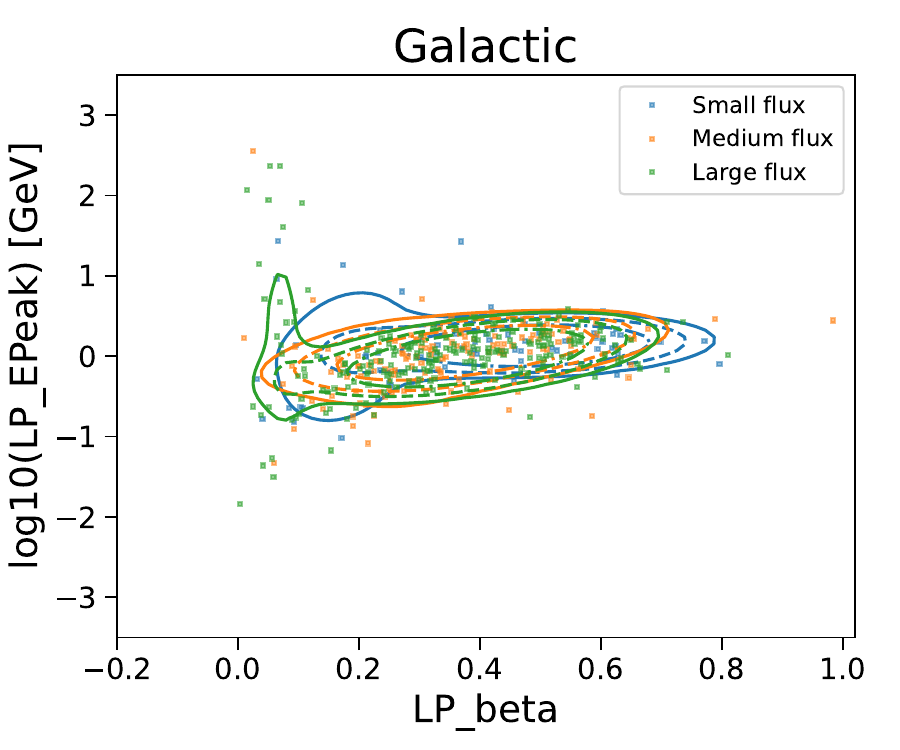}
\vspace{-3mm}
\caption{Distribution of associated and unassociated sources as well as extragalactic and Galactic sources for different fluxes.
The small, medium, and large flux bins are determined by the following boundaries in 
log10(Energy\_Flux100 [${\rm erg\,cm^{-2}\,s^{-1}}$]): (-13, -11.5, -11, -9).
}
\vspace{-4mm}
\label{fig:shift}
\end{figure}

In this paper, we use the 4FGL-DR4 catalog, version ``gll\_psc\_v34.fit''
\cite{2022ApJS..260...53A, 2023arXiv230712546B}.
An example of a dataset shift is shown in the top two plots in Fig.~\ref{fig:shift}.
The distribution of sources in the training dataset (associated sources) on the top left plot is different from the distribution of 
target dataset on the top right plot (the unassociated sources).
This difference can be due either to covariate shift or to prior shift.
In this note we perform classification of unassociated sources using both covariate and prior shift assumptions 
and discuss the relation between the two approaches.

\section{Data selection}
\label{sect:data}

In the analysis we use a similar definitions of classes as in Ref.~\cite{2024arXiv240104565M}.
The physical classes are grouped into four sets dominated by FSRQs, BL Lacs, pulsars, and millisecond pulsars: 
``fsrq+'': fsrq, nlsy1, css;
``bll+'': bll, sey, sbg, agn, ssrq, rdg;
``psr+'': snr, hmb, nov, pwn, psr, gc;
``msp+'': msp, lmb, glc, gal, sfr, bin
(the definitions of class acronyms can be fround in Ref.~\cite{2022ApJS..260...53A}).
We note that in these classes we do not take into account bcu and spp sources.
We use seven input features in the covariate shift case~\cite{2024arXiv240104565M}:
log10(Energy\_Flux100), log10(Unc\_Energy\_Flux100), log10(Signif\_Avg), LP\_index1GeV, LP\_beta, LP\_SigCurv, log10(Variability\_Index), where LP\_index1GeV is the index of the log-parabola spectral fit at 1 GeV, while the other features are transformations of the features in the 4FGL-DR4 catalog.
The classification is performed with the random forest algorithm implemented in 
scikit-learn \citep{scikit-learn}.
For the prior shift model we use the following three features: log10(Energy\_Flux100), LP\_beta, log10(LP\_EPeak), where 
LP\_EPeak is the peak energy of the spectral energy distribution modeled with the log-parabola function.
We consider only sources with 10 MeV $<$ LP\_EPeak $<$ 1 TeV.

\section{Prior shift model}
\label{sect:prior_shift}

In the prior shift model, we assume that the distribution of sources in different classes as functions of input features is the same for 
associated and unassociated sources $p_{\rm assoc} (x|k) = p_{\rm unas} (x|k)$,
while the difference in the distributions of  associated and unassociated sources comes from the differences in class prevalences.
The overall probability distribution function of unassociated sources is represented as:
\be
p_{\rm unas} (x) = \sum_k p_{\rm assoc} (x|k) \pi_k,
\ee
where $\pi_k$ is the frequency (prevalence) of class $k$ among unassociated sources.
The unknown coefficients $\pi_k$ are determined by maximizing the log-likelihood
\be
\log L = 
\sum_{i \in {\rm unas}} \log(p_{\rm unas} (x_i))
- N_{\rm unas} \int p_{\rm unas} (x) dx.
\ee

\begin{figure}[ht]
\centering
\vspace{-3mm}
\includegraphics[width={\size}cm,clip]{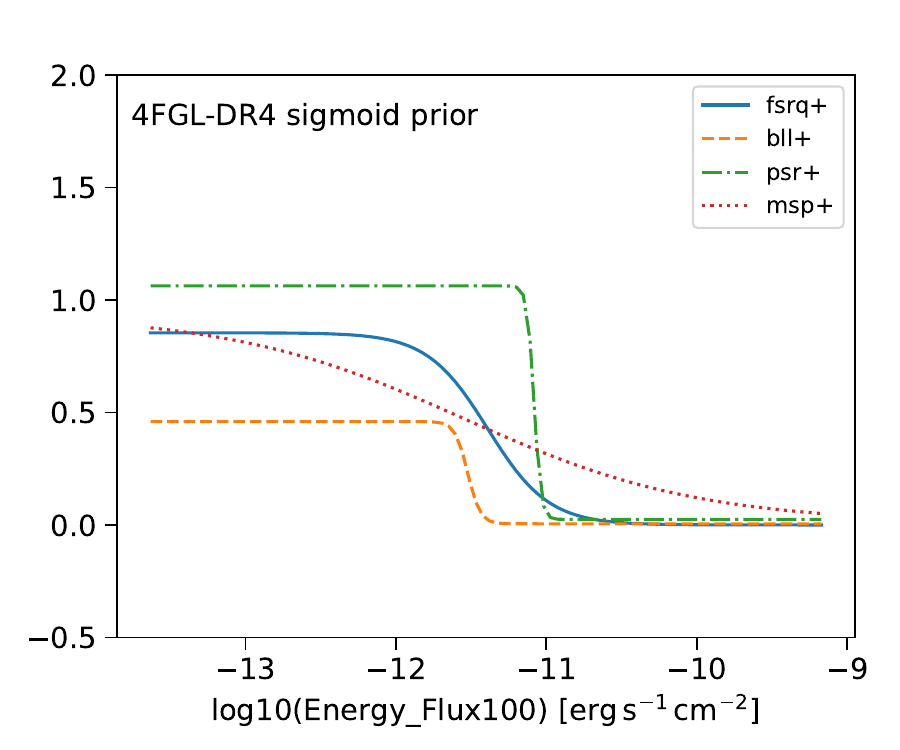}
\includegraphics[width={\size}cm,clip]{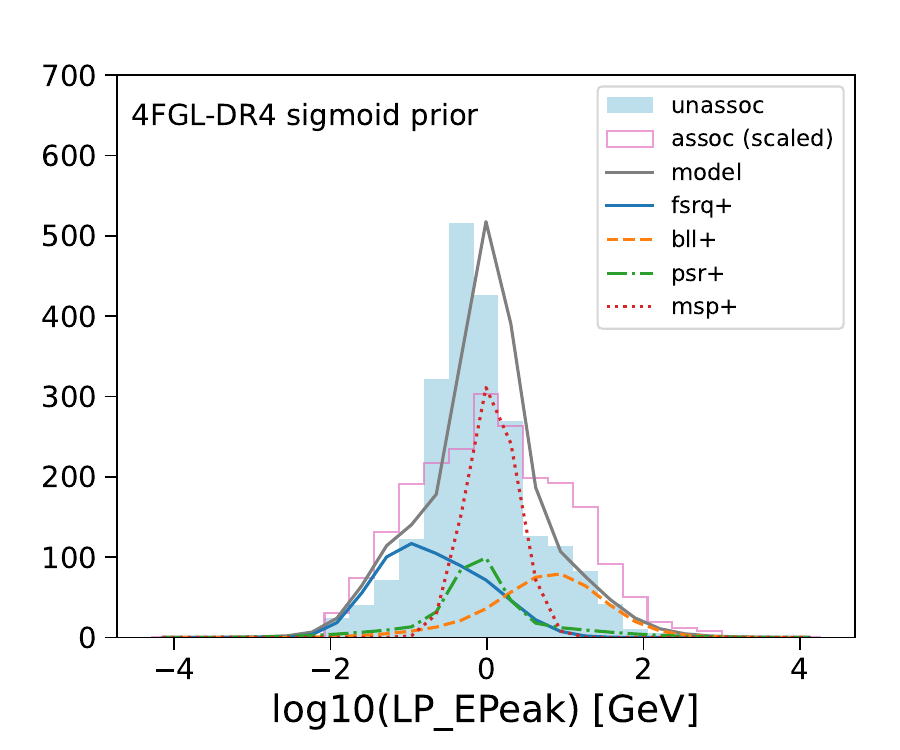}\\
\includegraphics[width={\size}cm,clip]{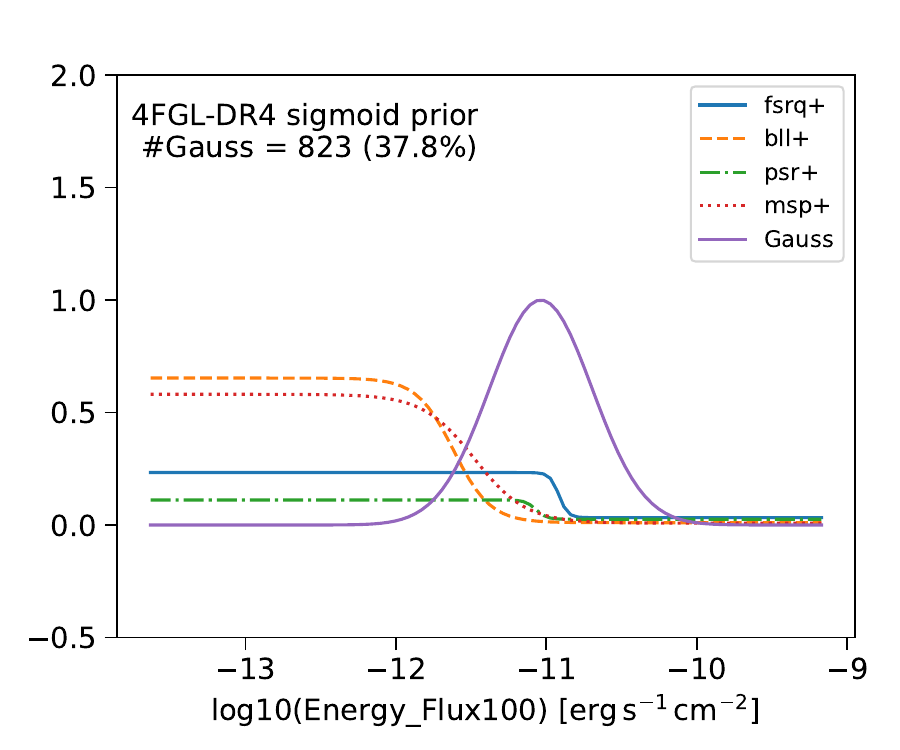}
\includegraphics[width={\size}cm,clip]{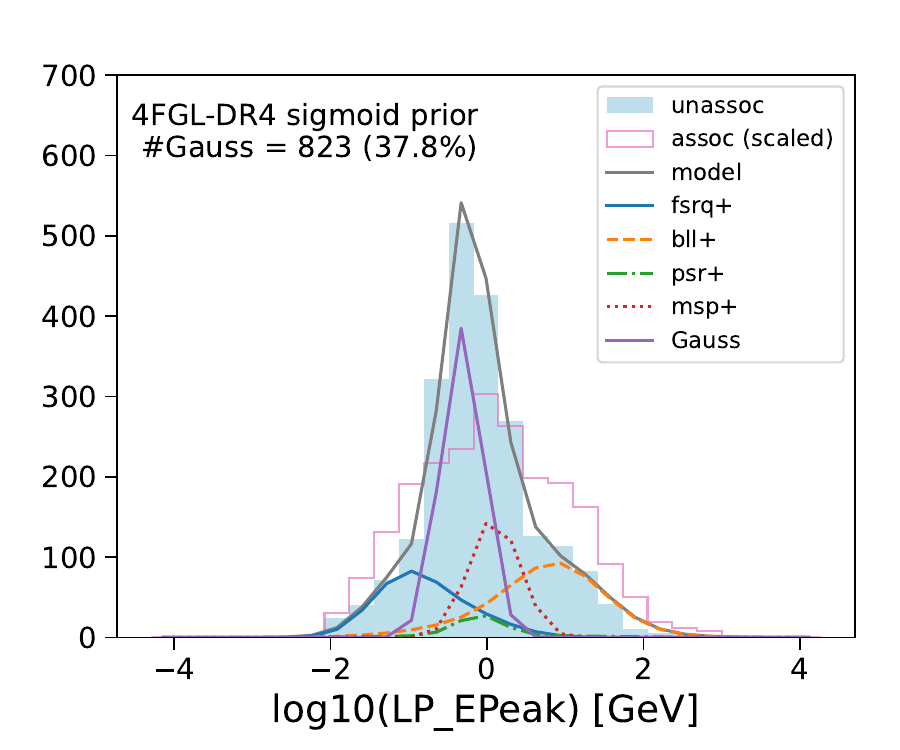}
\vspace{-3mm}
\caption{Flux-dependent prior shift models, cf. Eq.~(\ref{eq:pshift}). Top (bottom) panels: without (with) an additional Gaussian component. On the bottom-left panel we also show the Gaussian model PDF (normalized to one at maximum).}
\vspace{-3mm}
\label{fig:fdep}
\end{figure}

One of the caveats of the classical prior shift model is that the distributions may not be the same
for associated and unassociated sources in the different classes.
For example, the distribution of extragalactic associated sources in Fig.~\ref{fig:shift} bottom left has a large density at $\beta \approx 0$ for 
intermediate and large fluxes, while for small fluxes the distribution in $\beta$ is rather wide.
The top right plot of Fig.~\ref{fig:shift} shows that in the high LP\_EPeak regime (largely extragalactic from the bottom left plot), nearly all unassociated sources are at low fluxes.
%It is clear from the distribution of unassociated sources (top right panel of Fig.~\ref{fig:shift}) that, for example,
%the fraction of low-flux extragalactic sources is larger for unassociated sources compared to the fraction of low-flux extragalactic sources among the associated sources.
In order to account for the possible flux dependence of the distributions of sources, we introduce flux-dependent prior shifts, parameterized by sigmoid functions plus a constant:
\be
\label{eq:pshift}
\pi_k(x) = \frac{a}{1 + e^{(x - b) / c}} + d,
\ee
where $x =$ log10(Energy\_Flux100). This model has four (instead of one) free parameters for each of the classes.
The flux-dependent prior shifts are shown in Fig.~\ref{fig:fdep} top left panel.
As expected, the contribution of extragalactic sources (fsrq+ and bll+ classes) is suppressed at large fluxes.
However, the model does not fit the data well, as one can see, e.g., in the example of the log10(LP\_EPeak) distribution on the top right panel of 
Fig.~\ref{fig:fdep}.

One of the advantages of the prior shift model, is that it allows one to introduce new classes.
In addition to the flux-dependent prior shifts, we introduce a new population modeled as a 3-dimensional Gaussian distribution in the 3 input features of the prior shift model.
The flux-dependent prior shifts are shown on the bottom left panel of Fig.~\ref{fig:fdep}.
The corresponding model and the distribution of sources as a function of log10(LP\_EPeak) are shown on the bottom right panel of Fig.~\ref{fig:fdep}.
Now the model fits the data relatively well at the expense of a new Gaussian component parameterized with 7 parameters.

\section{Prior vs covariate shift models}
\label{sect:prior_vs_cov}

\begin{figure}[ht]
\centering
\vspace{-3mm}
\includegraphics[width={\size}cm,clip]{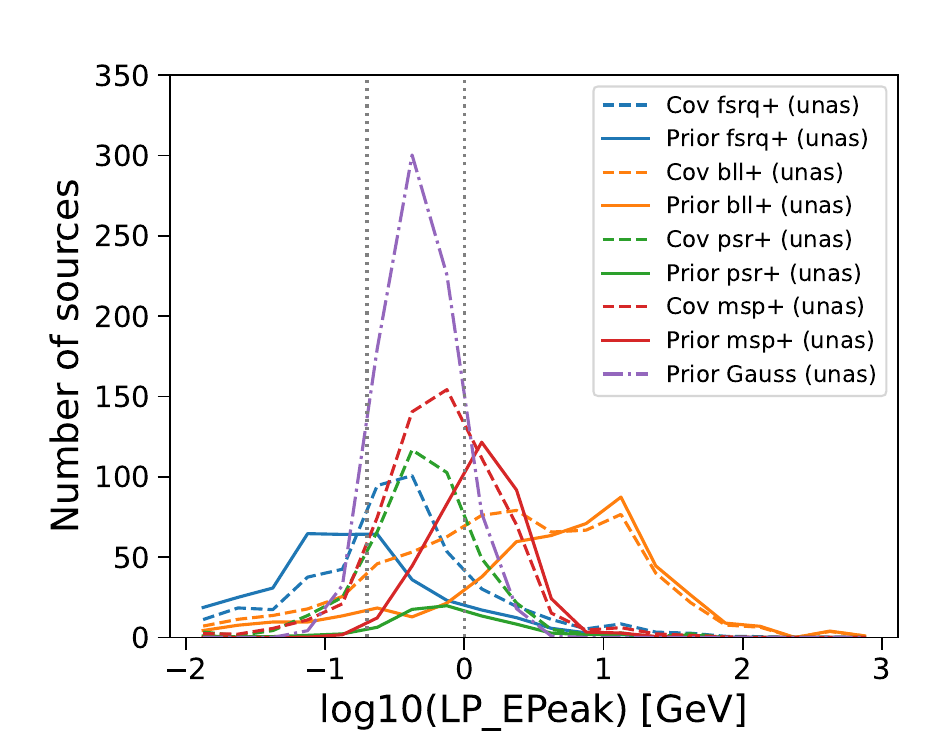}
\includegraphics[width={\size}cm,clip]{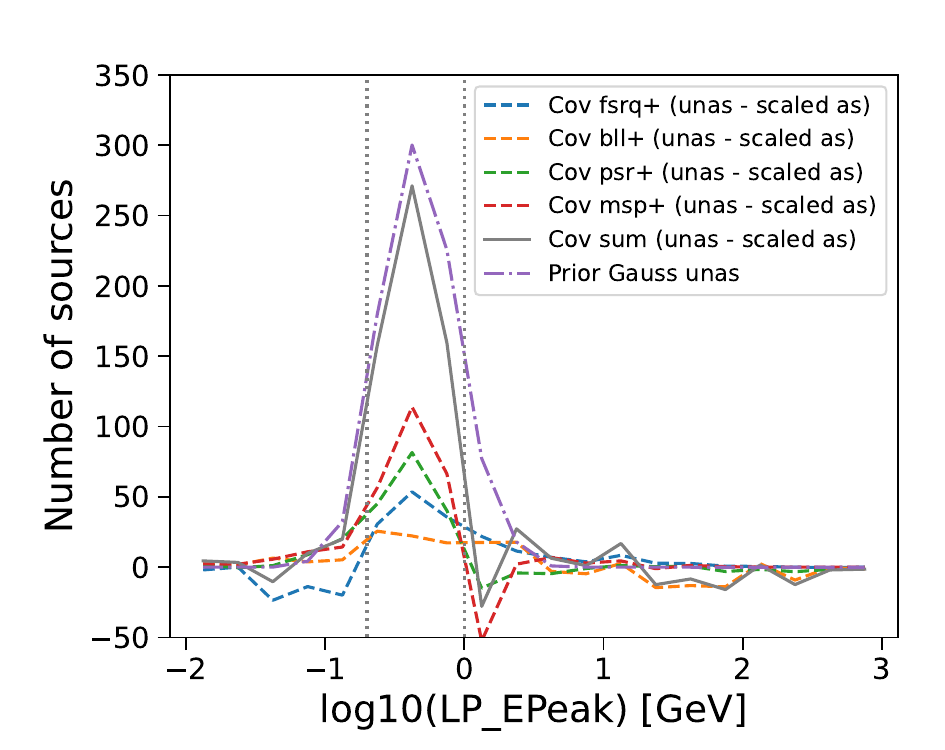}
\vspace{-3mm}
\caption{Left panel: comparison of prior and covariate shift models. Right panel: comparison of the sum of differences of covariate shift models minus scaled distributions of associated sources with the Gaussian component in the prior shift model.}
\label{fig:pr_vs_cov}
\vspace{-4mm}
\end{figure}

Although covariate shift models cannot accommodate a new class, one can indirectly see the presence of a possible new population of sources.
In the region of parameter space, where the new population may be present, the members of the new population would be proportionally distributed among the known classes. 
In Fig.~\ref{fig:pr_vs_cov} on the left, we show a comparison of the predictions for the distribution of the classes in the prior shift and covariate shift models as a function of log10(LP\_EPeak).
The models agree relatively well outside of the grey vertical lines at 200 MeV and 1 GeV, which approximately show the ``boundaries'' of the Gaussian component.
However, inside the vertical grey dotted lines the covariate shift model systematically predicts more sources for the four classes compared to the prior shift model. This effect can be due to a new population of sources modeled as a Gaussian in the prior shift case.
In order to qualitatively asses the contribution of the new component to the four classes in the covariate shift case, we scale the distributions of associated sources to fit the distributions of unassociated sources outside of the grey dotted lines and subtract the scaled distributions of associated sources from the unassociated ones.
The corresponding differences are shown as dashed lines on the right panel of Fig.~\ref{fig:pr_vs_cov}.
The sum of the differences is shown as the grey solid line. 
It has a similar distribution as the distribution of sources in the Gaussian component shown by the purple dash-dotted line.
This similarity shows that there is possibly a new population of sources modeled with a Gaussian distribution in the prior shift case. Given the relatively low Epeak values, this population should have a big overlap with the population of soft Galactic unassociated sources introduced in Ref.~\cite{2022ApJS..260...53A}.
%\medskip

%\vspace{-1mm}
{\bf Acknowledgments.}
The author would like to thank Jean Ballet, Aakash Bhat, Toby Burnett, and Benoit Lott for valuable discussions and comments and
to acknowledge support by the DFG grant MA 8279/3-1.
%and the use of the following software: Astropy (\url{http://www.astropy.org}) \cite{2013A&A...558A..33A},
%Matplotlib (\url{https://matplotlib.org/}) \cite{Hunter:2007}, 
%pandas (\url{https://pandas.pydata.org/}) \cite{mckinney-proc-scipy-2010},
%scikit-learn (\url{https://scikit-learn.org/stable/}) \cite{scikit-learn},
%and TensorFlow (\url{https://www.tensorflow.org/}) \cite{tensorflow2015-whitepaper}.

%
% BibTeX or Biber users please use (the style is already called in the class, ensure that the "woc.bst" style is in your local directory)
% \bibliography{your_bib_file} % Replace "your_bib_file" with the actual name of your .bib file
%
\vspace{-4mm}
\bibliography{prior_cov_ricap}

\end{document}